\documentclass[runningheads]{llncs}
\usepackage{graphicx}
\usepackage{tabularx}
\usepackage{makecell} 
\usepackage{bchart}
\usepackage{pgfplots}

\usepackage{float}

\begin{document}
\title{Trends in Machine Learning and Electroencephalogram (EEG): A Review for Undergraduate Researchers }
\titlerunning{Trends in Machine Learning and EEG}
%
\author{Nathan Koome Murungi\inst{1}\orcidID{0009-0008-3170-7689} \and
Michael Vinh Pham\inst{1}\orcidID{0009-0007-5322-1543}
\and
Xufeng Dai\inst{2}\orcidID{0009-0006-0513-1255}
\and
Xiaodong Qu\inst{1}\orcidID{0000-0001-7610-6475}\thanks{Nathan, Michael, and Xufeng are the first three authors of this paper, and they contributed equally. Professor Xiaodong Qu is the mentor for this research project.}}
\authorrunning{N. Murungi et al.}
%
\institute{Swarthmore College, Swarthmore PA 19081, USA 
\email{\{nmurung1,mpham1,xqu1\}@swarthmore.edu}
\and
Haverford College, Haverford, PA 19041, USA
\email{xdai1@haverford.edu}
}
\maketitle              
\begin{abstract}
This paper presents a systematic literature review on Brain-Computer Interfaces (BCIs) in the context of Machine Learning. Our focus is on Electroencephalography (EEG) research, highlighting the latest trends as of 2023. The objective is to provide undergraduate researchers with an accessible overview of the BCI field, covering tasks, algorithms, and datasets. By synthesizing recent findings, our aim is to offer a fundamental understanding of BCI research, identifying promising avenues for future investigations.

\end{abstract}
\section{Introduction}

Since the advent of computing, the disparity between human and computer technology has significantly diminished. Starting with early human-computer interfaces like keyboards and microphones, the boundary between humans and computers has been progressively blurred, primarily owing to the emergence and utilization of brain-computer interfaces \cite{padfield2019eeg}. In the rapidly advancing field of Brain-Computer Interfaces (BCI), Electroencephalography (EEG) analysis plays a crucial role in establishing a connection between the human brain and Machine Learning (ML) algorithms \cite{craik2019deep,gong2021deep,qu2022eeg4home,qu2020using,roy2019deep,saeidi2021neural,sha2020knn}. The proliferation of ML algorithms and the increasing availability of EEG data have created exciting opportunities for researchers to explore new approaches to interpreting raw EEG data. However, this progress presents a challenge for newcomers due to the overwhelming volume of research papers and the rapid rate at which they become outdated, making it challenging to navigate the research landscape effectively.

To tackle this formidable challenge, we present a meticulous examination of the existing literature in the field of Brain-Computer Interfaces (BCI), with a specific focus on the most recent advancements up to 2023. This paper is tailored to cater to undergraduate researchers, serving as a comprehensive overview and guide for those aspiring to conduct research in the Electroencephalography (EEG) domain. By systematically analyzing and organizing the obtained findings, our objective is to facilitate a profound comprehension of the current landscape of BCI research while identifying promising avenues for future investigations.

In addition to providing a comprehensive review, this paper specifically delves into utilizing Transformers, a prominent and rapidly emerging machine learning algorithm, within the realm of BCI research. By focusing on the application of Transformers in this context, we aim to shed light on its significance and impact, offering insights into its potential advantages and limitations. Table \ref{tab:acronyms} lists the acronyms used in this paper.

\begin{table} [b!]
\begin{center}
\begin{tabular}{ | c | c |}
\hline
 Abbreviation & Definition \\ [0.5ex] 
 \hline\hline
ML & Machine Learning\\
\hline
DL & Deep Learning\\
\hline
LSTM & Long Short-Term Memory\\
\hline
CNN & Convolutional Neural Network\\
\hline
DNN & Deep Neural Network\\
\hline
AE & Autoencoder\\
\hline
GAN & Generative Adversarial Network\\
\hline
SVM & Support Vector Machine\\
\hline
RNN & Recurrent Neural Network\\
\hline
ANN & Artificial Neural Network\\
\hline
RF & Random Forest\\
\hline
KNN & K-Nearest Neighbor\\
\hline
DBN & Deep Belief Network\\
\hline
EEG & Electroencephalography\\
\hline
BCI & Brain-Computer Interfaces\\
\hline
\end{tabular}
\end{center}
\caption{List of Acronyms}
\label{tab:acronyms}
\end{table}

\subsection{Research Questions} Our research aims to address the following questions at the intersection of ML and EEG research:
\begin{itemize}
\item What are the most suitable tasks, datasets, and ML algorithms for undergraduate researchers to explore the realms of Machine Learning (ML) and Electroencephalography (EEG)?

\item What are the prevailing trends observed within the intersection of ML and EEG in 2023?
\end{itemize}

By answering the first question, we aim to provide guidance to undergraduate researchers, helping them identify appropriate starting points and resources for their ML and EEG investigations. This will enable a seamless entry into the research domain and facilitate their understanding of fundamental concepts and methodologies.

Regarding the second question, we plan to provide an overview of the current trends within the EEG-ML field, highlighting the latest advancements and emerging research directions as of 2023. Understanding these trends may help undergrad researchers stay informed about cutting-edge developments and identify promising avenues for their investigations.

\section{Methods}

\begin{figure}[b!]
  \centering
  \includegraphics[width=\linewidth]{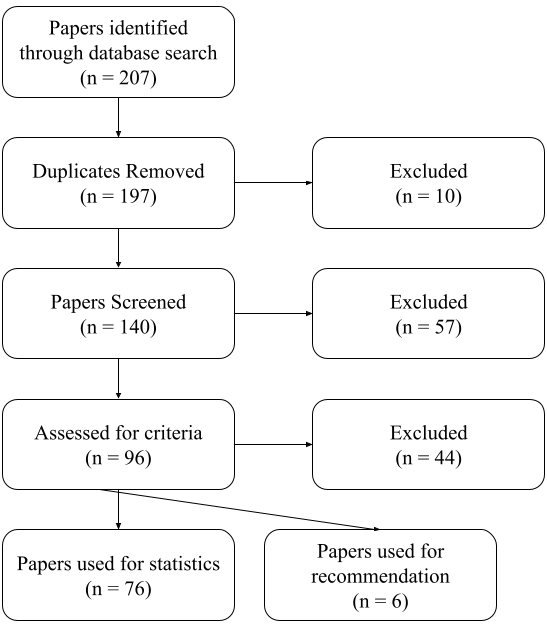}     
  \caption{Selection process for the papers}
  \label{fig:method}
\end{figure}

\subsection{Keywords}   Preferred Reporting Items for Systematic Reviews and Meta-Analyses (PRISMA) is the systematic review method that was used to identify relevant EEG research papers that use ML methods, as mentioned in \cite{craik2019deep,roy2019deep}. This search was conducted in February 2023
within Google Scholar, Paperwithcode, arXiv, and PubMed databases
using keywords: (\textit{'Machine Learning' OR 'LSTM' OR 'Deep Learning' OR 'CNN' OR 'RNN' OR 'Classification' OR 'DNN' OR 'Autoencoder' OR 'GAN' OR ' Transformer' AND 'Brain Computer Interface' OR 'Motor Imagery Classification' OR 'seizure' OR 'EEG Review' OR 'Speech' OR 'Emotion' OR  'Parkinson's' OR 'Alzheimer's' OR 'Depression' OR 'Rehabilitation' OR 'Gender Classification' OR 'Stroke' OR 'Person Identification' OR 'Age Classification' OR 'Task classification') AND  ('EEG' OR 'Electroencephalography' AND 'Survey' or 'Review'}). Figure \ref{fig:method} visualizes this screening method and shows how the best 76 papers were selected and further narrowed down to the 9 recommended papers for undergraduates, as they are limited in time.  Furthermore, with the understanding that undergraduate researchers are limited in time, we also narrowed all the papers we reviewed down to 9 suggested research papers that a new undergraduate researcher should start with and read to get acclimated with the current trends within the field.

The criteria for selecting appropriate papers are as follows: 
\begin{itemize}
    \item \textbf{EEG only: } We only look at Electroencephalography (EEG) studies with human subjects to keep type of data consistent.
    \item \textbf{Time:} Due to constantly new discoveries within the Machine Learning field and EEG field, we only included up-to-date literature published in 2020 and after.
    \item \textbf{ML/DL included:} Highly focused on Machine Learning and Deep Learning approaches to EEG processing.
    \item \textbf{Reproducibility:} Papers that include data preprocessing, feature extraction, results, code, and data source. 
     \item \textbf{Target Audience:} Comprehensible to undergraduate students majoring or minoring in computer science, interested in the topic of machine learning.
\end{itemize}

\section{Results}

\subsection{Tasks}

In the healthcare field, EEG data can be decoded to understand the underlying and psychological status of affected individuals\cite{altaheri2021deep,qu2020identifying}.

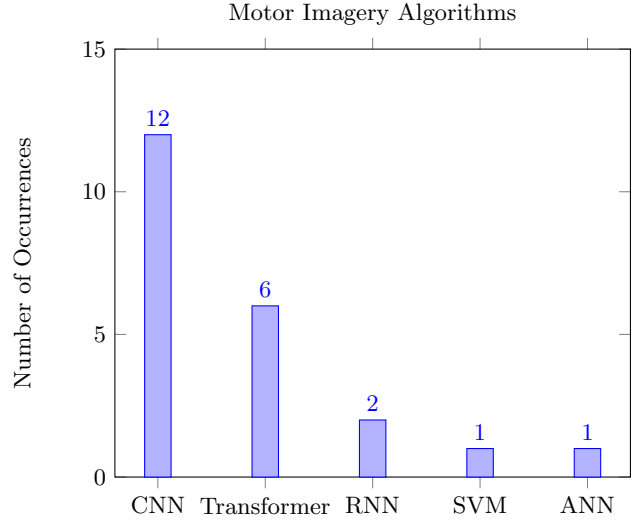
\begin{figure} [b!]
\centering
\begin{tikzpicture}
\begin{axis}[
    ybar,
    ymin=0,
    ymax=15,
    ylabel=Number of Occurrences,
    symbolic x coords={CNN,Transformer, RNN,SVM,ANN},
    xtick=data,
    nodes near coords,
    nodes near coords align={vertical},
    title={Motor Imagery Algorithms},
    ]
\addplot coordinates {(CNN,12) (RNN,2) (Transformer,6) (SVM,1) (ANN,1)};
\end{axis}
\end{tikzpicture}
\caption{Breakdown of Algorithms used for Motor Imagery Classification }
\label{fig:mi}
\end{figure}

From our review, most BCI research has been conducted in the clinical setting, specifically MI, Seizure, and Emotion Detection. As such, and due to their high accuracy, it's recommended that undergraduates start with these tasks. Table \ref{tab:taskcount} shows the most common tasks found for BCI research (with the top 3 bolded). Also there are nonclinical tasks, such as \cite{qu2018eeg,wang2022eeg,zhou2022brainactivity1} Figure \ref{fig:mi} shows the most common algorithms used for ML. The most common algorithms for Emotion were CNN, RNN, SVM, and KNN. For Seizure, the most commonly used were CNN, RNN, Transformer, and KNN. These top 3 tasks found in our literature review are described below.

\subsubsection{Motor Imagery}
Motor imagery EEG tasks involve the measurement and analysis of EEG brain activity patterns associated with mentally envisioning particular motor actions. Rehabilitation and assistive technologies can benefit from EEG/ML Motor Imagery by facilitating the development of solutions for individuals with motor and limb impairments via a means of communication from acquired neural activity of the kinesthetic imagination of limbs. \cite{singh2021comprehensive} Some healthcare systems utilize EEG signals to allow individuals to sense and interact with the physical world by controlling exoskeletons, wheelchairs, and other assistive technologies \cite{altaheri2021deep}. 

Not only can patients with motor disabilities benefit from EEG/ML Motor Imagery, but also everyday users through the development of Brain-Computer Interfaces to control specific motor actions. Current work done in classifying imaginations of the right or left hand and fingers for example could translate to a future where users are able to imagine specific motor actions such as moving a cursor, controlling a robotic arm, or navigating a video game. \cite{altaheri2021deep}

\subsubsection{Emotion Recognition}
Emotion detection EEG/ML tasks entail identifying and classifying patterns of EEG brain activity associated with different emotional states\cite{zhang2023multimodal}. Emotion is commonly associated with perception, decision-making, and human interaction. As the BCI field continues to grow, the interest in establishing an "emotional" interaction between humans and computers has increased in consumer products such as virtual reality \cite{suhaimi2020eeg}. In addition, a persons mental condition and emotional state can be apparent through their EEG waves. Therefore, emotion detection finds application in mental health contexts, enabling remote assessment and monitoring of patients' emotional states by therapists or counselors which can facilitate timely interventions and support and better treatment. Emotion recognition is the leading scientific problem in Affective Computing, which is how computer systems recognize and comprehend emotional information for natural human-computer interactions. With better Affective Computing, AI chatbots and voice assistants can better understand human users' emotions  providing more personalized and empathetic interactions. \cite{li2022eeg}

\subsubsection{Seizure Detection}

Epileptic seizures are chronic neurological diseases that can substantially impact the lifestyle of an affected individual. For some patients, there could be hundreds of seizures a day, which greatly affects their brain.  They are sudden abnormalities in the brain's electrical activities, which is manifested as excessive discharge of the neuronal network in the cerebral cortex \cite{ahmad2022eeg}. Accurate and timely detection for seizures can greatly help improve the livelihood of many affected individuals by preemptively describing seizure risk which can allow an individual to seek medical expertise. For studies that highlight seizure detection, EEG signals are recorded during the phases of a seizure: prodromal, early ictal, ictal, and postictal as well as seizure free periods. In addition, the control class consisted of non-epileptic subjects.

\begin{table} [t]
\begin{center}
\begin{tabular}{ | c | c |  }
\hline
 Task & Paper Count \\ [0.5ex] 
 \hline\hline
\textbf{Motor Imagery} & \textbf{19} \\ 
\hline
\textbf{Seizure} & \textbf{12} \\
\hline
\textbf{Emotion} & \textbf{12} \\
\hline
Parkinson's  & 3 \\
\hline
Gender/Age & 2 \\
\hline
Depression & 2 \\
\hline
Stroke & 1 \\
\hline
Person Identification & 1 \\
\hline
Inner Speech & 1 \\
\hline
Dementia & 1   \\
\hline
Lie Detection &  1   \\
\hline

\end{tabular}

\end{center}

\caption{Task Breakdown for Non-Review Papers}
\label{tab:taskcount}

\end{table}

\subsection{Algorithms}

Compared with review papers in previous years \cite{craik2019deep,dou2022time,qu2020multi,roy2019deep,zhao2020sea,zhao2019bira}, many of the algorithm findings were the same such as CNN dominating as the most frequently used in BCI research. But there were also interesting trends with Transformers and DBN. Transformers, a new ML algorithm, replace RNNs with sequential data because they have a self-attention mechanism that allows them to process input sequences of varying lengths and extract relevant information from them \cite{yi2022attention}. DBN also makes this list as an algorithm that has fallen out of trend. Despite making up a sizeable portion of earlier BCI research and review, rarely any research, we came across used DBN. Table \ref{tab:algocount} shows the most common algorithms found for BCI research.

\begin{table} [t!]
\begin{center}
\begin{tabular}{ | c | c | }
\hline
 Algorithm & Paper Count\\ [0.5ex] 
 \hline\hline
\textbf{CNN} & 20 \\
\hline
\textbf{RNN} & 13  \\
\hline
\textbf{Transformer} & 5 \\
\hline
SVM & 4  \\
\hline
KNN & 4 \\
\hline
ANN & 4 \\
\hline
RF & 3 \\
\hline
AE & 2 \\
\hline

\end{tabular}
\end{center}
\caption{Algorithm Breakdown for Non-Review Papers}
\label{tab:algocount}
\end{table}





\subsection{Datasets}

The datasets used in the studies vary based on the task that the study focuses on, and the most used are BCI Competition Dataset IV for validating signal processing and classification method; DEAP for analyzing human affect states; SEED, a various use EEG dataset; and EEGEyeNet, and Eye-tracking Dataset and Benchmark for Eye Movement Prediction. Other available datasets include ‘Thinking out loud,’ an EEG-based BCI dataset for inner speech recognition, ‘Feeling Emotions,’ and EEG Brainwave Dataset for emotion recognition. The datasets cover a wide range of topics for BCI research and are accessible to undergraduate students. 
\begin{table} [t!]

\begin{center}
\begin{tabular}{ | c | c | c | c | }
\hline
 Dataset & Task & Year & Cited \\ [0.5ex] 
 \hline\hline
 DEAP \cite{5871728} & Emotion & 2011  &  3439\\
\hline
BCI Competition IV \cite{tangermann2012review}& MI & 2012  &  783\\
\hline

Dreamer \cite{7887697} & Emotion  & 2016 & 517 \\
\hline

SEED \cite{6695876} & Emotion  & 2013 & 358 \\
\hline

Bonn \cite{li2013feature} & Seizure  & 2013 & 316 \\
\hline

CHB-MIT \cite{prasanna2021automated} & Seizure  & 2021 & 21 \\
\hline

EEGEyeNet \cite{kastrati2021eegeyenet} & MI  & 2021 & 16 \\
\hline



\end{tabular}

\end{center}
\caption{Dataset Breakdown for Non-Review Papers}
\label{tab:data}

\end{table}










\section{Discussion}
We identified the most recent trend in Machine learning BCI research based on the papers we reviewed. We presented our recommendations for the top three tasks, algorithms, datasets, and review papers for undergraduate researchers.

It is clear that transformers are becoming more and more utilized in the last year to classify EEG data. Transformers have traditionally been used in natural language processing (NLP) due to their ability to handle long-range dependencies \cite{sun2021eeg}. Therefore, transformers can be efficient with the recognition of EEG data, a set of time series signals. However, transformers are commonly paired with algorithms such as CNNs, RNNs, or DBNs. We further expand our analysis by exploring a range of scholarly articles that employ analogous time-series data \cite{basaklar2021wearable,chen2021data,deb2022systematic,deb2021trends,huang2022evaluation,jiang2022deep,liu2023applying,lu2023machine,lu2022cot,luo2022multisource,ma2022traffic,ma2020statistical,peng2022energy,shen2021semi,tang2023active,zhang2022biotic,zhang2021implementation,zhang2022attention,zong2022beta}.

\subsection{Guide of Techniques Required for Research}

In order for an undergraduate research to be successful in their research, knowledge of some techniques is required. These techniques include an understanding of a computer programming language such as Python and  various tools and packages.



\subsubsection{Tools and Packages}
Fluency in Python is extremely beneficial for replicating the works mentioned in this paper effectively. Its readability, position as the de facto language for ML, and extensive community support make it an excellent choice for beginners in the EEG-ML field. NumPy and Pandas are useful ML libraries for numerical operations and data manipulation, making them convenient for tasks such as data preprocessing and exploration for EEG data. Overwhelmingly, the aforementioned works employed the Python library Sci-kit Learn to train classical ML algorithms such as Random Forests and SVMs. Deeper machine learning algorithms such as CNNs and Transformers had split usage between the PyTorch and Tensorflow/Keras Python libraries. Amazon Web Services (AWS), a cloud computing solution, offers services such as SageMaker which offers a complete ML development environment and EC2, an elastic virtual server service, which enables the ability to launch and oversee virtual machine instances to run ML code.

In summary, Python is quite mandatory for understanding EEG-ML source code; Although not mandatory, familiarity and being comfirtable using Numpy, Pandas, Sci-kit Learn, PyTorch or Tensorflow/Keras, are advantageous in further enhancing the replication process. Finally, AWS or other cloud computing services are nice tools to have especially when working with large EEG datasets.

\subsubsection{Recommended Courses}

It is optional to enroll in a college course to learn Python. Many universities such as Stanford, Cornell, and MIT provide free online courses that are very efficient in their programming methods. Reading Python's online documentation is an alternative to gaining an proficiency of Python and its syntax. Researchers lacking ML knowledge can also use online courses to gain an understanding of the topic. Based on our own undergraduate research experience, we recommend that undergraduate computer science researchers watch Stanford CS 229 and CS 230 lectures online.

In addition, researchers should be comfortable utilizing NumPy, Pandas, Sci-kit Learn, PyTorch, and Tensorflow/Keras libraries. We recommend that researchers read each library's documentation which outlines the capabilities of each method in the libraries. These documentations can be found for free online.

\subsubsection{Recommended Papers} In addition to completing courses to understand the techniques and packages needed to get started in BCI machine learning research, we recommended that undergraduates read the papers outlined in table \ref{tab:recstart}. 

\begin{table} [t!]
\begin{center}
\begin{tabular}{ | c | c | }
\hline
Recommended Paper Title & Paper\\ [0.5ex] 
 \hline\hline
Deep learning-based electroencephalography analysis: a systematic review & \cite{roy2019deep} \\ 
\hline
Deep learning for electroencephalogram (EEG)
classification tasks: a review& \cite{craik2019deep} \\ 
\hline
 Fear of the CURE: A Beginner’s Guide to Overcoming Barriers in Creating \\a Course-Based Undergraduate Research Experience& \cite{govindan2020fear} \\
\hline
\end{tabular}
\caption{Recommended Papers to Start Research}
\label{tab:recstart}
\end{center}
\end{table}

\subsection{Algorithms}

\subsubsection{Frameworks for Recommended Algorithms}

CNNs are widely used in computer vision and image classifications. In \cite{liu2020deep}, they discuss a novel network that combines CNNs, SAE, and DNN to convert EEG time series data into 2D images for good emotion classification. Once the EEG data is transformed into 2D images, features extracted from the original EEG data are sent to the CNN. The model includes many pooling layers and one output layer. After pooling, data is subsampled into images with smaller sizes. The weights and filters are learned through back-propagation. \cite{liu2020deep}

SVMs are a very useful supervised machine learning algorithm. It works by separating the dataset into two sections. This sectioning of the data can either be linear or nonlinear. Linear sectioning includes a discriminant hyperplane, while nonlinear separation includes a kernel function. Though computational complexity is low for this algorithm, its performance is heavily dictated by the kernel function. \cite{sha2020knn}

Transformers are a trending algorithm for EEG classification. A transformer network is a neural network that is based on a self-attention mechanism. The network consists of encoders and decoders. Each encoder has two sub-layers, a multi-head self-attention mechanism, and a position-wise connected forward feeding network. There is a residual connection around the sub-layers, which is followed by a normalization layer. \cite{siddhad2022efficacy}

\subsubsection{Recommended Tasks} 

The tasks in Table \ref{tab:taskrec} are recommended for undergraduate researchers to include in their studies because they are extensively researched and have ML high accuracy. The paper count column counts the number of papers that include the use of transformers. 

As shown in table \ref{tab:taskrec}, transformers can be used on a majority of Brain-Computer Interface tasks, but are most commonly used for Motor Imagery, Emotion Recognition, and Seizure Detection.

\begin{table} [t!]
\begin{center}
\begin{tabular}{ | c | c | c |}
\hline
 Task & Recommended Paper & Paper Count \\ [0.5ex] 
 \hline\hline
\textbf{Motor Imagery} & \cite{li2020novel} & 6 \\ 
\hline
\textbf{Seizure} & \cite{qureshi2021machine} & 5\\ 
\hline
\textbf{Emotion} &  \cite{wang2021review} & 1\\
\hline
\end{tabular}
\caption{Recommended Tasks}
\label{tab:taskrec}
\end{center}
\end{table}

\subsubsection{Recommended Algorithms}
The algorithms in Table \ref{tab:algorec} are recommended. CNN is the most prominent; SVM is low in computation cost, high in accuracy, and easy to understand. Transformers are trending in BCI research, and have high accuracy and parallelization.

\begin{table} [t!]
\begin{center}
\begin{tabular}{ | c | c | c | }
\hline
 Algorithm & Recommended Paper \\ [0.5ex] 
 \hline\hline
\textbf{CNN} & \cite{liu2020deep} \\ 
\hline
\textbf{SVM} & \cite{sha2020knn} \\ 
\hline
\textbf{Transformer} &  \cite{siddhad2022efficacy} \\
\hline
\end{tabular}
\caption{Recommended Algorithms}
\label{tab:algorec}
\end{center}
\end{table}

\subsubsection{Forms of Transformers}According to our finds, transformers are commonly accompanied with other algorithms or are utilized in different forms. Table \ref{tab:combinationtransformer} outlines some of the forms of transformers used from the papers found.

\begin{table} [t!]
\begin{center}
\begin{tabular}{ | c | c | c | }
\hline
Algorithm  & Dataset &  Paper \\ [0.5ex] 
 \hline\hline
CNN + Transformer &  SEED & \cite{guo2022transformer}\\
\hline
DNN + Transformer& TUEG & \cite{kostas2021bendr}\\
\hline
S3T & BCI Competition IV  & \cite{song2021transformer}\\
\hline
\end{tabular}

\end{center}
\caption{Combination of Algorithms with Transformers}
\label{tab:combinationtransformer}
\end{table}

For several emotion recognition tasks, transformers were combined with CNNs.  CNNs are fed the input DE features and extract complete information for multi-frequency data with a depthwise convolution layer, which is then fed to a transformer model. \cite{guo2022transformer} 

A combination of DNNs and Transformers was for a Motor Imagery task. DNNs were used for pre-training, while transformers were used as the classification algorithm. \cite{kostas2021bendr}. 

S3T is a Spatial-Temporal Tiny Transformer used for Motor Imagery tasks due to its ability to capture spatial and temporal features. S3T requires an attention mechanism to transform the data into a highly distinguishing representation. The model consists of a spatial filter, spatial transformation, temporal transformation, and a classifier. \cite{song2021transformer}

\subsection{Datasets}

\subsubsection{Dataset Users}

Each dataset utilizes different subjects with the goal to be classified for different tasks. The goal of the BCI Competition datasets is to validate signal processing and classification methods for Brain-Computer Interfaces (BCIs). Each data set consists of single-trials of spontaneous brain activity, one part labeled (calibration or training data) and another part unlabeled (evaluation or test data), and a performance measure. The goal is to infer labels (or their probabilities) for the evaluation data sets from calibration data that maximize the performance measure for the true (but to the competitors unknown) labels of the evaluation data.

The DEAP dataset is a multimodal dataset for the analysis of human affective states. The EEG and peripheral physiological signals of 32 participants were recorded as each watched 40 one-minute long excerpts of music videos. Participants rated each video in terms of the levels of arousal, valence, like/dislike, dominance and familiarity.

The SEED dataset contains subjects' EEG signals when they were watching films clips. The film clips are carefully selected so as to induce different types of emotion, which are positive, negative, and neutral ones.

The STEW dataset consists of raw EEG data from 48 subjects who participated in a multitasking workload experiment utilizing the SIMKAP multitasking test. The subjects’ brain activity at rest was also recorded before the test and is included as well.

\subsubsection{Recommended Datasets}

Table \ref{tab:datarec} overviews a recommended dataset based on the aforementioned recommended tasks in Table \ref{tab:data}.






\begin{table} [t!]
\begin{center}
\begin{tabular}{ | c | c | c |}
\hline
 Task & Recommended Dataset  \\ [0.5ex] 
 \hline\hline
 Motor Imagery & EEGEyeNet\\
\hline
Emotion Recognition  & DEAP\\
\hline
Seizure Detection & Bonn\\
\hline
\end{tabular}
\end{center}

\caption{Dataset Recommendations}
\label{tab:datarec}
\end{table}

\subsubsection{Dataset Breakdown For Transformers}From the papers found that utilize transformers, table \ref{tab:datasettransformer} shows the breakdown of the datasets that were used. The other category is a combination of privately collected datasets or uncommon individual sets. DEAP and SEED are available to the public.

\begin{table} [t!]
\begin{center}
\begin{tabular}{ | c | c | }
\hline
 Dataset & Paper Count\\ [0.5ex] 
 \hline\hline
\textbf{DEAP} & 5 \\ 
\hline
\textbf{SEED} & 2 \\ 
\hline
\textbf{Other} &  8 \\
\hline

\end{tabular}
\caption{Dataset breakdown for Transformer Algorithm}
\label{tab:datasettransformer}
\end{center}
\end{table}

\subsection{Papers}

\subsubsection{Paper Recommendation for Undergraduate Researchers}We recommend 3 individual research papers in Table \ref{tab:indvrec} and 3 review papers in Table \ref{tab:revrec} to any new researcher to get acclimated with trends in BCI research.

\begin{table} [t!]
\begin{center}
\begin{tabular}{ | c | c | c | }
\hline
 Paper & Algorithm & Task \\ [0.5ex] 
 \hline\hline
\cite{siddhad2022efficacy} & Transformer & Age/Gender\\ 
\hline
\cite{hassin2022identification} & RF & Parkinson's\\
\hline
\cite{liu2020deep} & CNN/AE & Emotion\\
\hline

\end{tabular}
\end{center}
\caption{Individual Papers}
\label{tab:indvrec}
\end{table}

There is limited available literature where EEG data has been used for gender and age classification. The recommended paper for Age/Gender classification gives a firm overview of the task as well as the algorithm. The paper discusses how a transformer network was used for raw EEG data classification. The utilization of Transformers in the classification aided in the achievement of start-of-the-art accuracy for both gender and age. \cite{siddhad2022efficacy} Similarly, the other two papers give a firm overview of their respective tasks, as well as a clear insight into the framework and architecture of the algorithm.

\begin{table} [t!]
\begin{center}
\begin{tabular}{ | c | c | c | }
\hline
 Paper & Algorithm & Task \\ [0.5ex] 
 \hline\hline
\cite{saeidi2021neural}  & ML & BCI\\
\hline
\cite{hossain2023status} & DL & BCI\\
\hline
\cite{houssein2022human} & ML & Emotion\\
\hline

\end{tabular}

\end{center}
\caption{Review Papers}
\label{tab:revrec}
\end{table}

Each of the papers highlighted in table \ref{tab:revrec} gives a strong overview of either Machine Learning or Deep learning for their respective tasks. We recommend that undergraduate researchers read these papers and get an even stronger understanding of the BCI research landscape.

\subsubsection{Paper Recommendation for Undergraduate BCI research using Transformers}We recommended 3 individual research papers in Table \ref{tab:indvrectransformer}. Each paper has a different task and dataset. Each paper clearly explains its algorithms framework, methodologies, and results. 

\begin{table} [t!]
\begin{center}
\begin{tabular}{ | c | c | c | }
\hline
 Paper & Task & Dataset(s) \\ [0.5ex] 
 \hline\hline
 \cite{sun2021eeg}& Motor Imagery & HysioNet\\ 
\hline
  \cite{luo2020eeg}&  Emotion Recognition & DEAP/SEED\\
\hline
 \cite{deng2023eeg}&  Seizure Detection & CHB-MIT\\
\hline

\end{tabular}

\end{center}
\caption{ Recommended Transformer Papers}
\label{tab:indvrectransformer}
\end{table}

\subsection{Challenges}

One challenge encountered in utilizing ML algorithms for EEG task classifications is their limited robustness. Often, these algorithms are trained on data obtained from a single individual or a small group of individuals. Consequently, when tested on data from a different individual who was not part of the training set, ML algorithms tend to exhibit subpar performance, requiring additional calibration and fine-tuning. To address this issue, a potential solution lies in employing data augmentation techniques, such as Generative Adversarial Networks (GANs). GANs can generate synthetic EEG data, augmenting the existing dataset and enhancing the generality and robustness of the models.

The practical utilization of brain waves for real-world task classifications poses another challenge. Primarily, EEG data collection predominantly occurs in controlled environments, which fails to capture the authentic real-world conditions and factors that impact brain waves, including sensory stimuli and varying levels of concentration. Consequently, ML algorithms often exhibit under-performance when tested in real-world scenarios compared to their performance on laboratory test data. It is imperative to undertake a comprehensive evaluation and understanding of these factors during the design phase to effectively integrate Brain-Computer Interfaces (BCIs) into real-world applications within authentic environments.

\subsection{Future Work}

The utilization of transformers in classification models for BCI research has demonstrated promising improvements in accuracy across a range of tasks, rendering them applicable in the BCI and medical fields. Subsequent studies can delve deeper into the analysis and exploration of transformers, expanding upon the groundwork established by this paper, with the aim of enhancing their performance by potentially mitigating computational requirements while further improving accuracy. Similar avenues for improvement exist for other algorithms, such as Recurrent Neural Networks (RNNs) and Convolutional Neural Networks (CNNs).

In future research, it is imperative to investigate and identify algorithms that exhibit the highest accuracy for specific tasks while also considering their computational efficiency. This will provide clarity on which algorithms are the most optimal in terms of both accuracy and computational requirements for each task, thus enabling informed decisions in selecting the most suitable algorithms for specific BCI applications.


\section{Conclusion}
In this paper, we have presented a comprehensive overview of essential tasks, algorithms, and datasets that serve as foundational components for undergraduate researchers embarking on EEG-based Brain-Computer Interface (BCI) research using machine learning (ML).

Based on our extensive analysis, we recommend focusing on prominent tasks such as Motor Imagery, Seizure Detection, and Emotion Classification. These tasks offer a wealth of available datasets and exhibit high accuracy classification rates. To assist researchers in getting started, we have identified specific datasets that align with each task, providing valuable resources to facilitate their investigations.

Furthermore, we suggest the incorporation of popular algorithms such as Convolutional Neural Networks (CNNs), Support Vector Machines (SVMs), and Transformers. These algorithms have consistently demonstrated remarkable accuracy in EEG classification tasks, making them reliable choices for further exploration. Specifically, we have highlighted the growing trend of Transformers in EEG-based BCI research, providing an overview of the most common task-dataset-algorithm combinations associated with this emerging approach.

By following our recommendations, undergraduate researchers can establish a solid foundation in the field, enabling them to confidently contribute to the rapidly evolving landscape of BCI research. We believe that this systematic overview and the suggested starting points will empower newcomers, facilitating their exploration of novel techniques and inspiring their contributions to advancements in BCI research.

As the field continues to progress, it is crucial to stay abreast of the latest developments and embrace new opportunities for innovation. We hope that this paper serves as a valuable resource, guiding researchers towards exciting avenues and inspiring future discoveries in the dynamic realm of EEG-based Brain-Computer Interfaces.

%
%
%
 \bibliographystyle{splncs04}
 \bibliography{BCI_review}

\end{document}